\documentclass[aps,prl,reprint,superscriptaddress]{revtex4-1}

\usepackage{graphicx}
\usepackage{amsmath}
\usepackage{amsthm}
\usepackage{color}
\usepackage{ulem}
\usepackage{xcolor}
\usepackage{verbatim}
\begin{document}

\title{Dissipation Layers in Rayleigh-B\'{e}nard Convection: A Unifying View}
\author{K. Petschel}
\email[]{klaus.petschel@uni-muenster.de}
\author{S. Stellmach}
\affiliation{Institut f\"ur Geophysik, Westf\"alische Wilhelms-Universit\"at M\"unster, D-48149 M\"unster, Germany}
\author{M. Wilczek}
\author{J. L\"ulff}
\affiliation{Institut f\"ur Theoretische Physik, Westf\"alische Wilhelms-Universit\"at M\"unster, D-48149 M\"unster, Germany}
\author{U. Hansen}
\affiliation{Institut f\"ur Geophysik, Westf\"alische Wilhelms-Universit\"at M\"unster, D-48149 M\"unster, Germany}

\date{\today}
\begin{abstract}
Boundary layers play an important role in controlling convective heat transfer. Their nature varies considerably between different application areas characterized by different boundary conditions, which hampers a uniform treatment. Here, we argue that, independent of boundary conditions, systematic dissipation measurements in Rayleigh-B\'enard convection capture the relevant near-wall structures. By means of direct numerical simulations with varying Prandtl numbers, we demonstrate that such dissipation layers share central characteristics with classical boundary layers , but, in contrast to the latter, can be extended naturally  to arbitrary boundary conditions. We validate our approach by explaining differences in scaling behavior observed for no-slip and stress-free boundaries, thus paving the way to an extension of scaling theories developed for laboratory convection to a broad class of natural systems. 
\end{abstract}
\maketitle
Buoyancy driven fluid flows are ubiquitous in nature. They stir the turbulence on the solar surface \cite{Brun2004}, power the zonal wind pattern observed in the atmospheres of giant planets \cite{Heimpel2005, Gastine2012}, are responsible for Earth's plate tectonics \cite{Mckenzie1974,Trompert1998}, cause atmospheric cloud formation \cite{Weidauer2010} and thunderstorms \cite{Fischer2011}, and occur in biological systems \cite{Ghorai2005, Kuznetsov2011}. A simplified analog of such situations, the so-called Rayleigh-B\'enard configuration consisting of a plane fluid layer heated from below, served as a cornerstone for the development of hydrodynamic stability theory \cite{Bernoff1994, Chandrasekhar1981, Bolton1985} and has become a paradigm for studies of convective turbulence.

The key quantities for characterizing convective systems are the resulting heat and momentum transport, typically expressed in terms of the nondimensional Nusselt number $\mbox{Nu}$ and Reynolds number $\mbox{Re}$. Much research over the past years has focused on the prediction of these transport characteristics, with the Grossmann-Lohse scaling theory \cite{Grossmann2000} and its extensions \cite{Grossmann2001, Grossmann2002, Grossmann2004} being among the most prominent examples. Beyond their significance in testing our theoretical understanding against experimental evidence, such scaling theories are crucial in estimating the heat transport in natural situations, for which the control parameter values often differ from those accessible in laboratory or numerical experiments by many orders of magnitude \cite{Chilla2012}. 

Modern scaling theories \cite{Cioni1997,Grossmann2000} emphasize the importance of both the thermal and the viscous boundary layer in controlling the scaling behavior. While revealing an impressive consistency with available experimental data \cite{Ahlers2009}, these theories unfortunately cannot be applied  directly to natural situations like the ones mentioned above. The problem is that the classical picture of a viscous boundary layer, in which the tangential velocity components decrease rapidly over a small, $O(\mbox{Re}^{1/2})$ length scale toward the boundary, is inextricably linked to the presence of a rigid boundary surface. In most of the examples mentioned above, however,  there is no rigid boundary forcing the tangential velocities to drop to zero, and consequently no classical viscous boundary layer is to be expected. Theories relying on the presence of such a layer, including the scaling theories mentioned earlier, therefore cannot be generalized to arbitrary boundary 
conditions in a straightforward manner.

In this Letter we introduce the concept of what we call dissipation layers (DL), a generalization of the classical boundary layers that is based on the kinetic energy budget and the thermal variance balance. To test this concept, we study these dissipation layers for both laboratory-style no-slip boundary conditions and so-called stress-free conditions, in which the horizontal shear stresses are required to vanish at the boundary. The classical boundary layer picture only applies to the former, but not to the latter case, in which the horizontal velocities tend to peak on the boundary itself. Interestingly, we find that pronounced dissipation layers are observed for both types of boundary conditions and in many respects behave qualitatively similar. This paves the way for applications of the existing theories to natural systems.

Since the boundary layer dynamics is controlled by molecular diffusion processes, it  is very sensitive to changes in Prandtl number $\mbox{Pr}=\nu/\kappa$, where $\kappa$ and $\nu$ are the thermal and momentum diffusivities. Furthermore, $\mbox{Pr}$ varies strongly for different fluids, ranging from $\mbox{Pr}=O(10^{-6})$ in stellar plasmas to $\mbox{Pr}=O(10^{23})$ in planetary mantles. Both facts have motivated several numerical studies \cite{ Kerr2000, Schmalzl2002, Breuer2004,Silano2010} on the influence of $\mbox{Pr}$ on convection. Surprisingly, for stress-free boundary conditions, scaling laws similar to the ones proposed by Grossmann and Lohse have been observed  \cite{Breuer2004}. Pointing out the absence of classical viscous boundary layers for stress-free boundaries, previous authors \cite{Breuer2004} thus concluded that these do not control the observed scaling and thus casted doubt on the universal applicability of the Grossmann-Lohse theory.

A number of important questions immediately arise in this context. (i) Is there, for stress-free conditions, a viscous layer close to the boundary \citep{Breuer2004,Eckhardt2001}
that plays a role similar to the classical viscous boundary layer in the no-slip case? (ii) If so, is there a unifying way of defining near-wall viscous layers independently of the exact nature of the mechanical boundary conditions? (iii) Can this unified definition also be extended to describe the thermal boundary layers? (iv) If this is the case, are the properties of these newly defined layers compatible with the basic assumptions of the established theories? (v) Finally, can these layers be connected to specific scaling regimes for $\mbox{Nu}$ and $\mbox{Re}$, effectively allowing a generalization of the existing theories?

Here, we take the view that the most basic feature of a boundary layer region is that molecular dissipation processes play a major role there, as has been demonstrated in previous work for laboratory-style, no-slip boundaries \citep{Stevens2010,Shishkina2010}. We thus focus on the horizontally averaged kinetic energy dissipation rate,
\begin{equation}
\left\langle \epsilon_u\right\rangle_h(z) = -2\mbox{Pr}\langle \operatorname{Tr}\left(S^2\right) \rangle_h \quad,
\end{equation}
defined by the trace of the squared rate-of-strain tensor $S$, and the corresponding thermal dissipation rate, 
\begin{equation}
\left\langle\epsilon_{T}\right\rangle_h(z)=\langle-\left(\nabla T\right)^2\rangle_h\quad,
\end{equation}
where the governing equations have been nondimensionalized by means of the system height, the temperature drop between the bottom and the top, and the thermal diffusion time. Using these quantities, the edges of the viscous and thermal dissipation layers are defined by the vertical positions where the local viscous and thermal dissipation rates equal their volume-averaged values, 
\begin{equation}
\left\langle \epsilon_u \right\rangle_h(z_{u,\mathrm{DL}})\stackrel{!}{=}\left\langle \epsilon_u\right\rangle_V  \; 
\text{and} \; \left\langle \epsilon_T \right\rangle_h(z_{T,\mathrm{DL}})\stackrel{!}{=}\left\langle \epsilon_T \right\rangle_V.
\end{equation}
The distances from the closest boundary then define the viscous and thermal dissipation layer thicknesses $\lambda_{u,\mathrm{DL}}$ and $\lambda_{T,\mathrm{DL}}$. 

Note that this definition, which separates the flow into regions of high (above average) and low (below average) thermal and kinetic energy dissipation, is independent of the nature of boundary conditions.
We also focus on dissipation rates here because these are expected to be related to the overall heat transport; in particular, we have \cite{Chilla2012}
\begin{equation}
  \label{eq:e_u}\left\langle \epsilon_u\right\rangle_V = (\mbox{Nu}-1)\mbox{Ra}\mbox{Pr} \quad \mbox{and} \quad
 \left\langle \epsilon_T\right\rangle_V = \mbox{Nu}\quad.
 \end{equation}

In the following, we investigate the properties of these dissipation layers in detail. Three-dimensional direct numerical simulations of  Rayleigh-B\'enard convection in a Boussinesq fluid with periodic boundary conditions in the horizontal direction have been carried out using an accurate pseudospectral method \cite{Stellmach2008}. The top and bottom boundaries are impermeable, kept at a fixed temperature, and either the horizontal velocities (no-slip) or the shear stress (stress-free) are assumed to vanish at the boundary. For a fixed Rayleigh number $\mbox{Ra}=5 \times 10^6$, the Prandtl number range  $0.01 \le \mbox{Pr} \le 300$ is systematically explored. Even for this moderate Rayleigh number, spatial resolutions up to $576^3$ grid points were necessary to adequately  resolve the Kolmogorov scales within the bulk at low $\mbox{Pr}$ \cite{Shishkina2010, Grotzbach1983}.

\begin{figure}
	\includegraphics[width=0.43\textwidth]{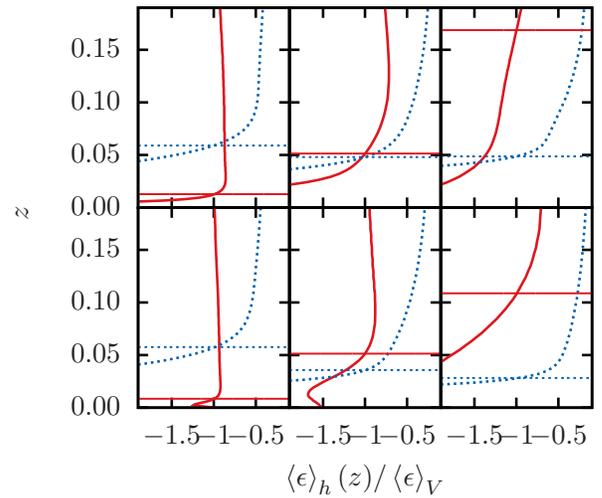} \caption{Temporally averaged depth profiles of the horizontally averaged viscous dissipation rate (red solid line) and the thermal dissipation rate (blue dashed line) for no-slip (upper graph) and stress-free (lower graph) boundaries and for different Prandtl numbers. The horizontal axis is scaled by the corresponding globally averaged dissipation rates [cf.~Eq.~\eqref{eq:e_u}]. The thickness of the newly defined dissipation layers is illustrated for each Prandtl number and boundary condition, respectively.} \label{fig:slice}
\end{figure}

\begin{figure}[hb]
	\includegraphics[width=0.4\textwidth]{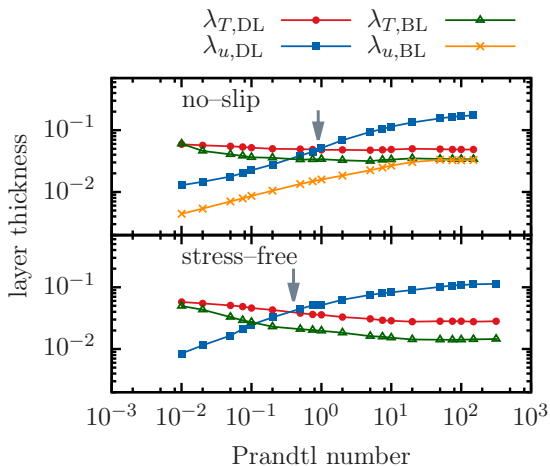} \caption{Thickness of the dissipation layers versus the Prandtl number for no-slip (upper graph) and stress-free (lower graph) boundaries. Red circles represent the thermal dissipation layer while blue squares denote the viscous dissipation layer. The gray arrows indicate the crossover of the thicknesses of both dissipation layers ($\lambda_{T,\mathrm{DL}}=\lambda_{u,\mathrm{DL}}$). The classical thickness definitions of the thermal and viscous boundary layer, $\lambda_{T,\mathrm{BL}}$ and $\lambda_{u,\mathrm{BL}}$, are denoted by green triangles and orange crosses, respectively. Note that for stress-free boundaries, the classical viscous boundary layer definition cannot be applied. \label{fig:layer_pr-scaling_t}}
\end{figure}

Figure \ref{fig:slice} shows the height-resolved dissipation rates for no-slip and stress-free boundaries and varying Prandtl numbers. All cases exhibit regions of strongly enhanced dissipation close to the boundary, validating the concept of dissipation layers separating the boundaries from the bulk. The profiles and consequently the layer thicknesses strongly depend on $\mbox{Pr}$: While for low Prandtl numbers the viscous layer is much smaller than the thermal layer, the opposite is observed for high Prandtl numbers. A crossover of the layer thicknesses takes place around $\mbox{Pr}=1$ for no-slip and slightly below for stress-free boundaries (cf.~also Fig.~\ref{fig:layer_pr-scaling_t}). This ``change of hierarchies'' separating the parameter space into two distinct regimes has been predicted for the no-slip case in previous theoretical works \cite{Tilgner1996, Grossmann2000}.

\begin{figure}[hb]
	\includegraphics[width=0.4\textwidth]{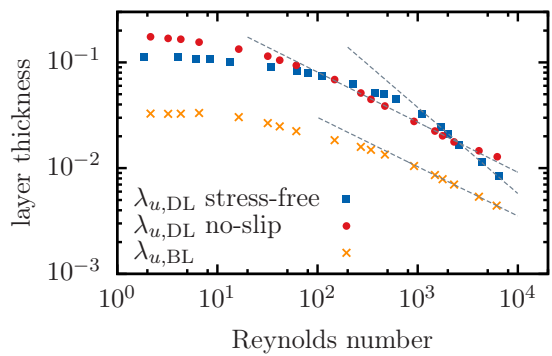} \caption{Thickness of different definitions of a viscous layer versus the Reynolds number $\mbox{Re}=\sqrt{\langle \boldsymbol{u}^2\rangle_V}/\mbox{Pr}$. The orange crosses denote a classical definition of a viscous boundary layer thickness $\lambda_{u,\mathrm{BL}}$, while the blue squares and the red circles denote the dissipation layer thickness $\lambda_{u,\mathrm{DL}}$  for stress-free and no-slip boundary conditions, respectively.}\label{fig:layer_re-scaling}
\end{figure}
The dissipation layer thicknesses along with their classical counterparts $\lambda_{T,\text{BL}}$ and $\lambda_{u,\text{BL}}$ are shown as a function of $\mbox{Pr}$ in Fig.~\ref{fig:layer_pr-scaling_t}. The classical boundary layer thicknesses are defined using linear fits of the depth profiles in the vicinity of the boundary, i.e. the so-called slope method \cite{Li2012}. It is observed that, apart from a constant prefactor in  amplitude, the viscous dissipation layer shows a similar Prandtl number dependence as the classical viscous boundary layer in the no-slip case: The layer thickness increases with $\mbox{Pr}$ and starts to saturate for the highest Prandtl numbers. The thermal dissipation layer and classical thermal boundary layer also show similar behavior and decline slowly with $\mbox{Pr}$. Differences appear for the lowest Prandtl number, which can be explained by the fact that the thermal boundary layer in this case is less well defined due to a finite thermal mean gradient in the bulk. We conclude that the newly defined viscous and thermal dissipation layers capture the near-wall characteristics of the system equally well as the classical approaches, which, however, fail to provide a concise definition for a viscous boundary layer in the stress-free case. 

\begin{figure}[ht]
	\includegraphics[width=0.4\textwidth]{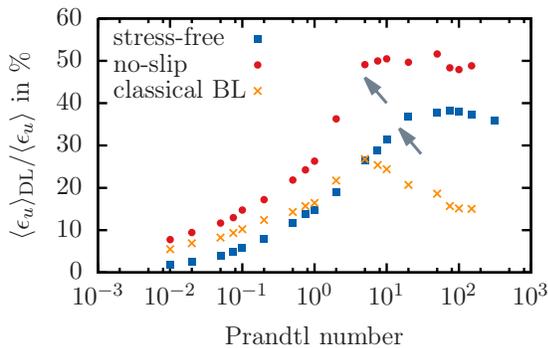} \caption{Ratio of the DL contribution of the globally averaged kinetic dissipation rate $\epsilon_u$ versus the Prandtl number. The blue squares denote results of simulations with stress-free boundaries, while red circles denote results with no-slip boundary conditions. The gray arrows divide the Prandtl number range into a bulk and a dissipation layer dominated regime. For completeness, the decomposition  employing the classical boundary layer definition is represented by orange crosses. } \label{fig:decomp-scaling}
\end{figure}
\begin{figure}[hb!]
	\includegraphics[width=0.4\textwidth]{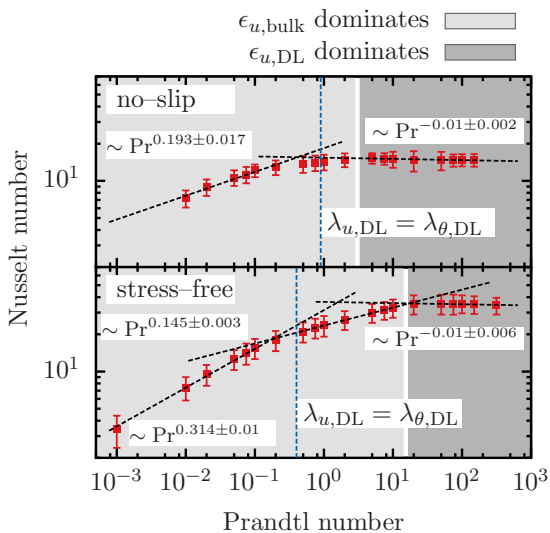} \caption{Nusselt number versus the Prandtl number for no-slip (upper graph) and stress-free (lower graph) boundaries. The vertical dashed line indicates the change of hierarchies, i.e., the Prandtl number where the thermal and the viscous dissipation layers are of the same thickness. The color-coded background represents the ratio of the DL contribution of $\epsilon_u$ in Fig.~\ref{fig:decomp-scaling}, where light gray indicates a bulk dominated regime and dark gray indicates a dissipation layer dominated regime. Data fits for the no-slip case are in good agreement with the predictions by Grossmann and Lohse \cite{Grossmann2001} ($\sim\mbox{Pr}^{1/5}$ for low Prandtl numbers)}\label{fig:nu-scaling}
\end{figure}

We now demonstrate that the new framework indeed allows a generalization to stress-free boundary conditions. As a central result we observe in Fig.~\ref{fig:layer_pr-scaling_t} that the layer thicknesses show a  functional form very similar to the no-slip case. The viscous dissipation layer thickness increases with $\mbox{Pr}$ with a tendency to saturate at high values, whereas the thermal dissipation layer thickness decreases with Prandtl number consistently with its classical counterpart. Again, the change of dissipation layer hierarchies is clearly visible for stress-free boundary conditions. 

The usual way to estimate the viscous boundary layer thickness is to apply the classical Prandtl-Blasius theory  to convective systems \cite{Zhou2010a, Zhou2010,Du_Puits2007,Li2012}, resulting in the prediction $\lambda_u\sim\mbox{Re}^{-1/2}$ \cite{Schlichting2000}. To check whether or not this prediction also holds for the proposed dissipation based thickness definitions, Fig.~\ref{fig:layer_re-scaling} shows $\lambda_{u,\mathrm{DL}}$ as a function of $\mbox{Re}$ for both stress-free and no-slip boundary conditions. For convenience, the classical viscous boundary layer thickness $\lambda_{u,\mathrm{BL}}$ is also shown for the no-slip case. Fits to the numerical data for large  $\mbox{Re}$ yield $\lambda_{u,\mathrm{BL}}\sim\mbox{Re}^{-0.466\pm0.004}$ and $\lambda_{u,\mathrm{DL}}\sim\mbox{Re}^{-0.474\pm 0.011}$ for the no-slip case, both in fair agreement with the classical Prandtl-Blasius prediction. In contrast, for the stress-free case, $\lambda_{u,\mathrm{DL}}$ decreases faster with $\mbox{Re}$, resulting in a scaling law $\lambda_{u,\mathrm{DL}}\sim\mbox{Re}^{-0.814\pm0.043}$. 

Within the parameter range covered by our study the system shows a transition from bulk dominated viscous dissipation at low Prandtl number to a regime characterized by significant dissipation within the dissipation layer at high Prandtl number. Figure \ref{fig:decomp-scaling} shows the ratio of the dissipation layer contribution $\langle\epsilon_u\rangle_{\mathrm{DL}}$ to the globally averaged kinetic dissipation rate $\langle\epsilon_u\rangle_V$ versus the Prandtl number. In the low Prandtl number regime, $\langle\epsilon_u\rangle_{\mathrm{DL}}/ \langle\epsilon_u\rangle_V$ is small and therefore a strong dominance of the bulk contribution is observed for stress-free and no-slip boundaries, respectively. The dissipation layer contributions then increase with Prandtl number, before saturating at  $\mbox{Pr}=O(10)$ with a contribution of approximately $50\%$ for the no-slip case and $35\%$ for the stress-free case. The saturation point marks a clear change in dynamical behavior, and, borrowing from the terminology of Grossmann and Lohse, we use it to quantify the transition from bulk to dissipation layer dominance.

On the basis of our newly defined dissipation layers, we can identify three different regimes within the investigated Prandtl number range, with the limits being defined by the crossover of the thermal and the viscous dissipation layer thicknesses and by the saturation of the dissipation layer contribution to $\left<\epsilon_u\right>_V$. In other words, the regimes are classified by the ``hierarchy'' of the viscous and thermal dissipation layers and by the relative strength of viscous bulk and near-wall dissipation. To illustrate that these regime transitions have dynamical relevance, Fig.~\ref{fig:nu-scaling} shows the Nusselt number obtained from numerical experiments versus the Prandtl number. In the stress-free case, we find three clearly distinguishable scaling regimes, each characterized by a different Nusselt-Prandtl scaling law.

Remarkably, the observed transitions between these scaling laws agree well with the regime limits described above. For no-slip boundaries, the intermediate Prandtl number regime appears to be very narrow. It is unclear whether the corresponding data points reflect a distinct power law, or merely represent a gradual transition between high and low Prandtl number scalings. This observation is consistent with the fact that our regime classification also suggests a very narrow intermediate Prandtl number regime. It is also consistent with the theory of Grossmann and Lohse, which, for the Rayleigh numbers used in this study,  predicts an intermediate regime so narrow that the authors are led to question its very existence in their original work \cite{Grossmann2000}. Future simulations at higher values of ${\mbox{Ra}}$ are needed in order to obtain more conclusive data. 

The results presented herein answer many of the questions posed at the beginning of this Letter. In particular, we have shown that near walls, narrow regions of enhanced viscous and thermal dissipation are a generic feature of turbulent convection irrespective of the exact nature of the mechanical boundary condition. This observation allows for a definition of thermal and viscous ``dissipation layers'' that neither relies on the applicability, nor on the validity of classical boundary layer models.  In cases where these dissipation layers have classical boundary layer counterparts, both layers have been shown to exhibit a similar Prandtl number scaling, albeit with a different prefactor. Different from their classical counterparts, however, the thermal and viscous dissipation layer thicknesses are shown to cross around $\mbox{Pr}=O(1)$ for both no-slip and stress-free boundary conditions. This separates the parameter space into distinct regions characterized by different dissipation layer hierarchies, in accordance with the theoretical predictions \cite{ Tilgner1996,Cioni1997,Grossmann2000}. The contribution of the dissipation layers to the overall dissipation has also been shown to be consistent with the theoretical assumptions. Finally, a regime classification based on dissipation layer hierarchies and on their contribution to the overall dissipation correlates well with transitions observed in the $\mbox{Nu}(\mbox{Pr})$ scaling. This strongly suggests that existing scaling theories developed for laboratory convection can indeed be extended to boundary conditions relevant for a broad class of natural systems.

An interesting finding in this context is that for stress-free boundary conditions, the viscous dissipation layer thickness decreases more rapidly with $\mbox{Re}$ than expected from a balance between horizontal advection and vertical diffusion,  which would lead to the classical $\lambda_{u,\mathrm{DL}} \sim \mbox{Re}^{-1/2}$ scaling. Although an explanation of this observation currently remains elusive, it might have strong implications for scaling transitions caused by the change in dissipation layer hierarchy. We might speculate that future studies of convection with stress-free boundaries are likely to reveal exciting surprises, such as scalings different from those predicted for no-slip systems by the current theories \cite{Schumacher2009, King2009}. 

\bibliography{/ddfs/user/data/k/k_pets02/Bibliography/Library/Library}

\end{document}